\documentstyle[prb,aps,epsfig,floats]{revtex}

\begin{document}
\draft
\preprint{}
%
\twocolumn[\hsize\textwidth\columnwidth\hsize\csname
@twocolumnfalse\endcsname
%

\title{Evolution of a Metal to Insulator Transition in 
Ca$_{2-x}$Na$_{x}$CuO$_{2}$Cl$_{2}$, as seen by ARPES}

\author{F. Ronning$^{1}$\cite{fr}, T. Sasagawa$^{2}$, Y. Kohsaka$^{2}$, 
K.M. Shen$^{1}$, A. Damascelli$^{1}$\cite{dlf}, 
C. Kim$^{1\delta}$\cite{cyk}, T. Yoshida$^{1}$, 
N.P. Armitage$^{1}$\cite{npa}, 
D.H. Lu$^{1}$, D.L. Feng$^{1}$\cite{dlf}, L.L. Miller$^{3}$, H. 
Takagi$^{2}$, Z.-X. Shen$^{1}$}
\address{$^{1}$
Department of Physics, Applied Physics and Stanford Synchrotron
Radiation Laboratory,\\ Stanford University, Stanford, CA 94305,
USA}

\address{$^{2}$Department of Advanced Materials Science, University of 
Tokyo, 
7-3-1 Hongo, Bunkyo-ku, Tokyo 113-0033, Japan}

\address{$^{3}$Department of Physics, Iowa State University, Ames Iowa, 
50011}

\date{\today}
\maketitle
\begin{abstract}

We present angle resolved photoemission (ARPES) data on Na-doped
Ca$_2$CuO$_2$Cl$_2$. We demonstrate that the chemical potential shifts
upon doping the system across the insulator to metal transition. The
resulting low energy spectra reveal a gap structure which
appears to deviate from the canonical d$_{x^{2}-y^{2}}$ $\propto |cos(k_x
a)-cos(k_y a)|$ form. To reconcile the measured gap structure
with {\it d}-wave superconductivity one can understand the data in terms
of two gaps, a very small one contributing to the nodal region and a very
large one dominating the anti-nodal region. The latter is a result of the
electronic structure observed in the undoped antiferromagnetic insulator.
Furthermore, the low energy electronic structure of the metallic sample
contains a two component structure in the nodal direction, and a change in
velocity of the dispersion in the nodal direction at roughly 50 meV. We
discuss these results in connection with photoemission data on other
cuprate systems.

\end{abstract}
%
\vskip2pc]
\narrowtext

\section{Introduction}

\begin{figure}[t!]
	\centering
	\includegraphics[width=3in]{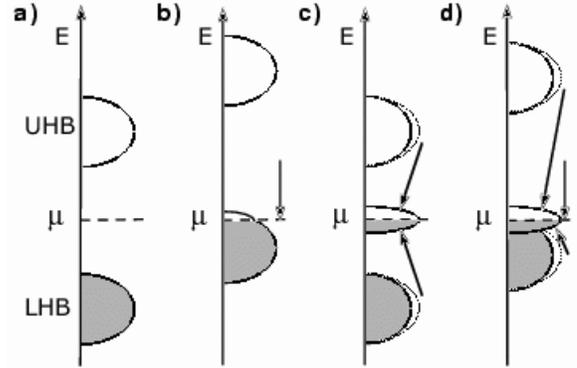}
	\vspace{0.1in}
\caption{Schematic doping evolution of an insulator. a) Upper (UHB) and 
lower 
(LHB) Hubbard bands for a 
	Mott Insulator. b) Result of doping holes into a band insulator. c) 
	Result of creating states inside the gap with doping. d) A scenario 
	combining the ideas of b) and c).}
	\label{mushiftcartoon}
\end{figure}

The close proximity of an antiferromagnetic Mott insulating phase suggests
that this insulator may be the natural starting point for understanding
the unusually high superconducting transition temperatures as well as the
unconventional normal state properties\cite{TimuskRPP99} in the cuprates.
The insulator to metal transition in strongly correlated materials is a
well studied fundamental problem in solid state physics in its own right.  
Even by limiting ourselves to a one band Hubbard model, which may be an
appropriate starting point for the half-filled
cuprates,\cite{AndersonScience87} there remain several ways one could
imagine the electronic structure evolving upon doping (see figure
\ref{mushiftcartoon}). One scenario is the chemical potential shifting to
the top of the valence band similar to the case of a band
insulator.\cite{MeindersPRB93} An alternative scenario is states being
created inside the Mott gap upon doping.\cite{AllenPRB90,InoPRB00} These
are the two most commonly discussed pictures with regard to 
the cuprates.

ARPES studies of the La$_{2-x}$Sr$_x$CuO$_{4}$ (LSCO) system show that
spectral weight is transfered to low energy states, which lie inside the
Mott gap, upon hole
doping.\cite{AllenPRB90,InoPRB00,RombergPRB90,ChenPRL91} In contrast, in
the Bi$_2$Sr$_2$Ca$_{1-x}$Y$_x$Cu$_2$O$_{8+\delta}$ (Bi2212) and
Nd$_{2-x}$Ce$_x$CuO$_4$ (NCCO) systems there is work which shows that the
chemical potential shifts in addition to a transfer of spectral
weight.\cite{VanVeenendalPRB94,HarimaPRB01} A potential problem for the
Bi2212 system is the difficulty to obtain deeply underdoped samples.  The
oxychlorides, Sr$_{2}$CuO$_{2}$Cl$_{2}$ and Ca$_{2}$CuO$_{2}$Cl$_{2}$, are
single layer cuprates that have provided the best photoemission data on
the undoped insulating state to date.  Thus the oxyhalides are ideal
systems to study the metal to insulator transition with photoemission, as
high quality single crystals, which cleave as easily as the Bi2212 system,
can now be grown through the metal to insulator transition.

Recently, we have reported that the chemical potential shifts upon doping
the insulator, Ca$_{2}$CuO$_{2}$Cl$_{2}$(CCOC), with Na.\cite{Kohsaka2001}
Here we present the full body of evidence which supports arguments for
such a shift of the chemical potential in the
Ca$_{2-x}$Na$_{x}$CuO$_{2}$Cl$_{2}$ system.  We examine the valence band
and photon energy dependence of Na-doped CCOC and compare it with that
found in the insulator.  By making a detailed comparison of the low energy
spectra we unambiguously show that the high energy pseudogap is in fact a
result of the {\it d}-wave-like dispersion seen in the insulator, as had
been suggested previously.\cite{LaughlinPRL97,RonningScience98}

As we will show, residual effects from the insulator strongly distort the
"band structure" in these deeply underdoped samples resulting in a gap
structure along the LDA-predicted Fermi surface that strongly deviates
from the canonical d$_{x^{2}-y^{2}}$ form. In order to reconcile this
result with {\it d}-wave superconductivity, we conclude that the measured
gap structure is a consequence of probing two different regimes. One
regime is concerned with a small unresolved gap along a hole pocket about
$(\pi/2,\pi/2)$. The second regime is in the vicinity of $(\pi,0)$ where a
large gap, which we identify as the pseudogap, is identified with the
undoped insulator. This picture is consistent with the observed chemical
potential shift.

The evolution of the electronic structure, however, is not solely a simple
shift of the chemical potential since the spectral lineshape also evolves
with doping.  We further investigate here the two component structure
found in the nodal direction of
Ca$_{2-x}$Na$_{x}$CuO$_{2}$Cl$_{2}$.\cite{Kohsaka2001} We show that this
structure is not directly related to T$_{c}$, and from a MDC (Momentum
Distribution Curve)  analysis we find a ``kink'' in the dispersion which
has the same energy as the "dip" which divides the two components in the
EDCs (Energy Distribution Curves). The observation of a kink near 50 meV
in this system supports the observed universality of this type of
behaviour in the cuprates.\cite{LanzaraNature} Finally, we discuss the
similarities and differences between photoemission results from other
cuprates and Na-doped CCOC.

\section{Experimental}

\begin{figure}[b!]
\centering 
\includegraphics[width=3in]{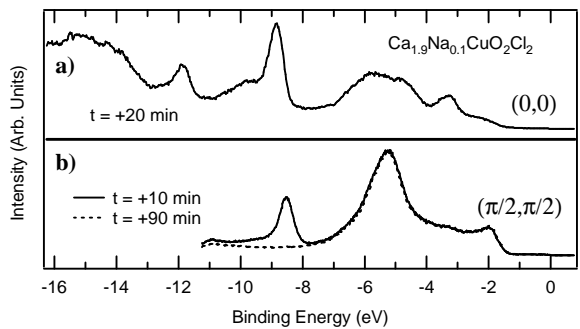}
\vspace{0.1in}
\caption[Time 
dependence of Ca$_{2-x}$Na$_{x}$CuO$_{2}$Cl$_{2}$ valence band 
spectra]{a) ARPES data reveals three features, at roughly 9, 10, and 12 
eV, 
in Na-doped CCOC not present in the pure sample.  The time elapsed between 
the initial cleave and the time the spectra were recorded is indicated.  
b) illustrates, with a different cleave, how the features 
vanish within 90 minutes of the initial cleave.  The photon energy was 
25.5 eV and the temperature was 20 K. The Fermi cutoff on the low energy 
spectral weight is too small to be visible in this intensity scale.}  
\label{FigNa5}
\end{figure}

Na-doped CCOC single crystals were grown by a flux method under high
pressure, by using a cubic anvil type pressure
apparatus.\cite{KohsakaGrowth} A powder mixture of Ca$_2$CuO$_2$Cl$_2$,
NaCl, NaClO$_4$, and CuO (1:0.2:0.2:0.1 molar ratio), sealed in a gold
tube, was heated to 1250$^{\circ}$C and then slowly cooled to
1050$^{\circ}$C under high pressure. Uniform samples at various Na
concentrations, x, were achieved by using different pressures during the
synthesis. 4 GPa and 5 GPa gave a Na concentration of 10$\%$ and 12$\%$,
respectively.  Crystals with a maximum dimension of 1.5 $\cdot$ 1.5
$\cdot$ 0.1 mm$^{3}$ were obtained.  The Na content was estimated by
comparing the c-axis lattice constant with powdered ceramic samples of
known content.\cite{Hiroi} Magnetization measurements of T$_{c}$ gave
results consistent with the assigned doping concentrations.  The Na
concentrations used in this study were x=0.10(T$_{c}$=13) and
x=0.12(T$_{c}$=22 K). Since superconductivity does not occur until
x$\sim$0.08 in the Na-doped CCOC system,\cite{KohsakaGrowth,Hiroi} we
consider our samples to lie in the heavily underdoped regime.  As a
reference, the optimal T$_{c}$ in powder samples has been found to be 28
K.\cite{Hiroi} ARPES measurements were performed on beamline 5-4 of SSRL.
The samples were easily cleaved in situ and measured at the indicated
temperatures in the figure captions. The angular resolution was
0.29$^{\circ}$, while the energy resolution was always $\leq$30 meV. When
examining the detailed low energy excitations (ie after figure
\ref{FigNature1}) the energy resolution was improved to $\leq$15 meV.

\section{Data and Results}
\subsection{Valence Band}

We begin with the valence band spectra of Na-doped CCOC in figure
\ref{FigNa5}, which illustrates a peculiarity of this system.  Namely,
from 8 to 13 eV binding energy there exists several pronounced features
which rapidly vanish with time.  The spectra from the cleave where these
features were most prominent is shown in the top panel.  Features are
visible at 9, 10, and 12 eV which are not observed in the parent compound,
CCOC. Figure \ref{FigNa5}b reveals that within 90 minutes following the
initial cleave, the additional features from 8 to 11 eV have vanished.  
The feature at 12 eV also vanishes (not shown).  For binding energies less
than 7 eV no change in the spectra was observed over these time scales
where particularly close attention was paid to the near $E_{F}$ features,
less than 1 eV binding energy. The photon energy was adjusted to ensure 
the
high binding energy features were not derived from a contribution of
second order light from the monochromator.  Comparing the spectra in
panels a) and b) suggests a dispersive nature of the additional peaks not
seen in the insulating compound.  However, it was also observed that these
features shift to lower binding energy as they age.  We tentatively assign
the origin of these features from 8 to 13 eV binding energy to flux
inclusions, such as NaCl and NaClO$_{4}$ which are used in excess during
the synthesis;\cite{BeardSSS94} subsequently the surface of these flux
inclusions are passivated by the photon following the cleave.  Despite the
uncertain nature of these high energy features, the fact that the low
binding energy region of the spectra have no variation over the short time
scale in which the high energy features vanish, gives us confidence that
our low energy results on the Na-doped CCOC crystals are intrinsic.

\begin{figure}[t!]
\centering 
\includegraphics[width=3in]{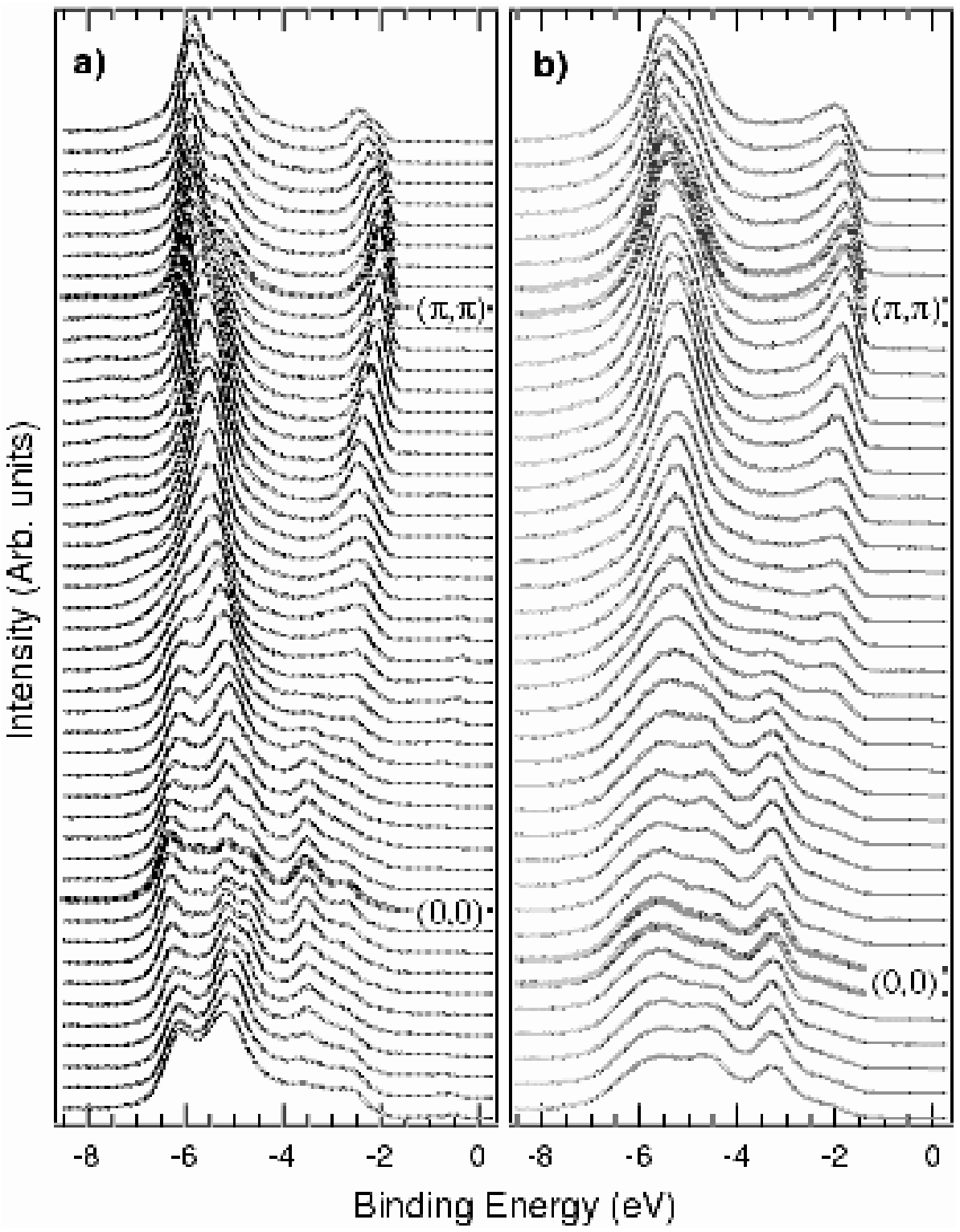}
\vspace{0.1in}
\caption[Valence band comparison of Ca$_{2}$CuO$_{2}$Cl$_{2}$ and 
Ca$_{1.9}$Na$_{0.1}$CuO$_{2}$Cl$_{2}$ along the nodal direction]
{Valence band spectra along the nodal 
direction for a) Ca$_{2}$CuO$_{2}$Cl$_{2}$ and b) 10$\%$ Na-doped 
CCOC. $E_{\gamma}$=25.5 eV, 
T=200 K and 20 K for x=0 and x=0.1, respectively.}
\label{MvsIVB}
\end{figure}

\begin{figure}[t!]
\centering 
\includegraphics[width=3in]{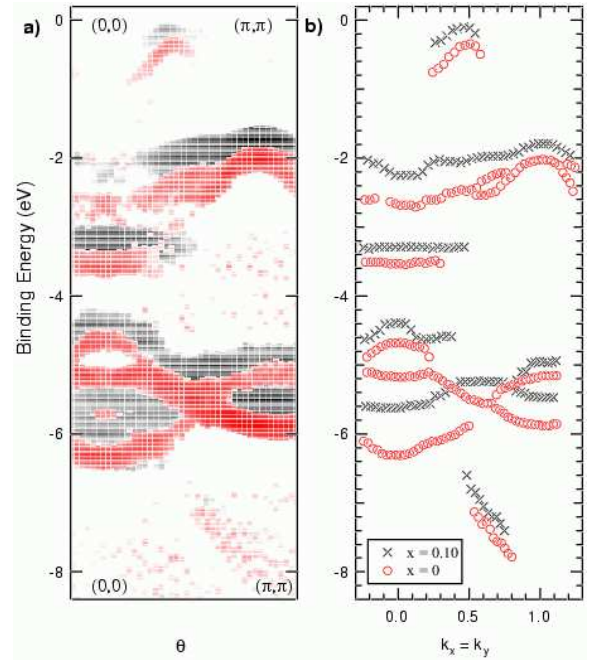}
\vspace{0.1in}
\caption[Valence band comparison of Ca$_{2}$CuO$_{2}$Cl$_{2}$ and 
Ca$_{1.9}$Na$_{0.1}$CuO$_{2}$Cl$_{2}$ along the nodal direction]
{(color) Valence band comparison of Ca$_{2}$CuO$_{2}$Cl$_{2}$ and 10$\%$ 
Na-doped 
CCOC.  a) is a second derivative plot of the EDCs presented in figure 
\ref{MvsIVB} retaining only 
points with negative curvature, and b) maps the dispersion of the various 
features as seen by eye.  The measurement conditions are 
identical except for the temperature of the insulator which is raised
to avoid electrostatic charging during the 
photoemission process. $E_{\gamma}$=25.5 eV, 
T=200 K and 20 K for x=0 and x=0.1, respectively.  The data are 
consistent with a shift of the chemical potential with doping as 
discussed in the text.}
\label{MvsIVBEk}
\end{figure}

Figure \ref{MvsIVB} compares the valence band spectra along the nodal
direction for x=0 and x=0.1.  The data on the insulator are consistent
with previous reports on a similar material
(Sr$_{2}$CuO$_{2}$Cl$_{2}$),\cite{DurrPRB00,PothuizenPRL97} and the data
for the Na-doped Ca$_{2}$CuO$_{2}$Cl$_{2}$ were reproduced on multiple
cleaves.  The position of the valence band in the insulator with respect
to the chemical potential can vary as much as 1 eV which is believed to be
a result of pinning the chemical potential at different impurity levels
that exist for the different cleaved surfaces. It is important to note
that the valence band data of the insulator shown in figures \ref{MvsIVB}
and \ref{MvsIVBEk} are at the minimum possible binding energy, since if
they are shifted further towards the chemical potential, spectral weight
from the tail of the valence band peaks would be pushed above the chemical
potential falsely indicative of a metallic nature. The low energy
excitations within 1 eV of the chemical potential, which the remainder of
this paper will focus on, are barely visible on this broad scale.  
Generally, the two data sets are quite similar, although differing
relative intensities for a few of the bands about (0,0) can be seen.  The
other difference is that the insulator data are sharper than those of the
metal.  Perhaps this is an indication that the surface of Na-doped
Ca$_{2}$CuO$_{2}$Cl$_{2}$ is not as well ordered as the pure sample grown
under atmospheric conditions.  Figure \ref{MvsIVBEk}a plots the second
derivative of the above spectra, which can be used to more clearly
identify the dispersion of various features.  This method agrees with the
dispersion extracted by the peak positions of the various bands in the
insulator and the metal as shown in panel \ref{MvsIVBEk}b.  The resolved
bands appear to have shifted to lower binding energy with doping by
roughly 300 meV, as would be expected if this were a simple band material.  
One apparent exception exists from the band at 5.6 eV binding energy in 
the
metallic sample near (0,0) in figure \ref{MvsIVBEk}b.  However, due to the
broadness of the features in the metal, it is likely that this band is a
superposition of several peaks, resulting in this apparent discrepancy.  
Note that in the insulator several bands are resolved in this same region.  
Thus, the overall picture from the valence band data is highly suggestive
of a chemical potential shift with doping.  The most convincing
demonstration for this is at $(\pi,\pi)$ where the peaks for both
concentrations are most clearly resolved.  Here, the peak positions
indicate that the insulator is shifted to higher energy by at least 300 
meV
relative to the metal.  The specific value will depend on which insulating
data set is used, since the chemical potential in the insulator can vary
substantially as mentioned above. We again emphasize that the current
binding energy of the insulator could not be found at smaller binding
energies without spectral weight being shifted unrealistically above the
chemical potential. Thus, neither charging of the insulator nor the
variability of the chemical potential in the insulator can be responsible
for the observed shift upon doping

\subsection{Near E$_{F}$ Comparison}

\begin{figure}[b!]
 	\centering
\includegraphics[width=3in]{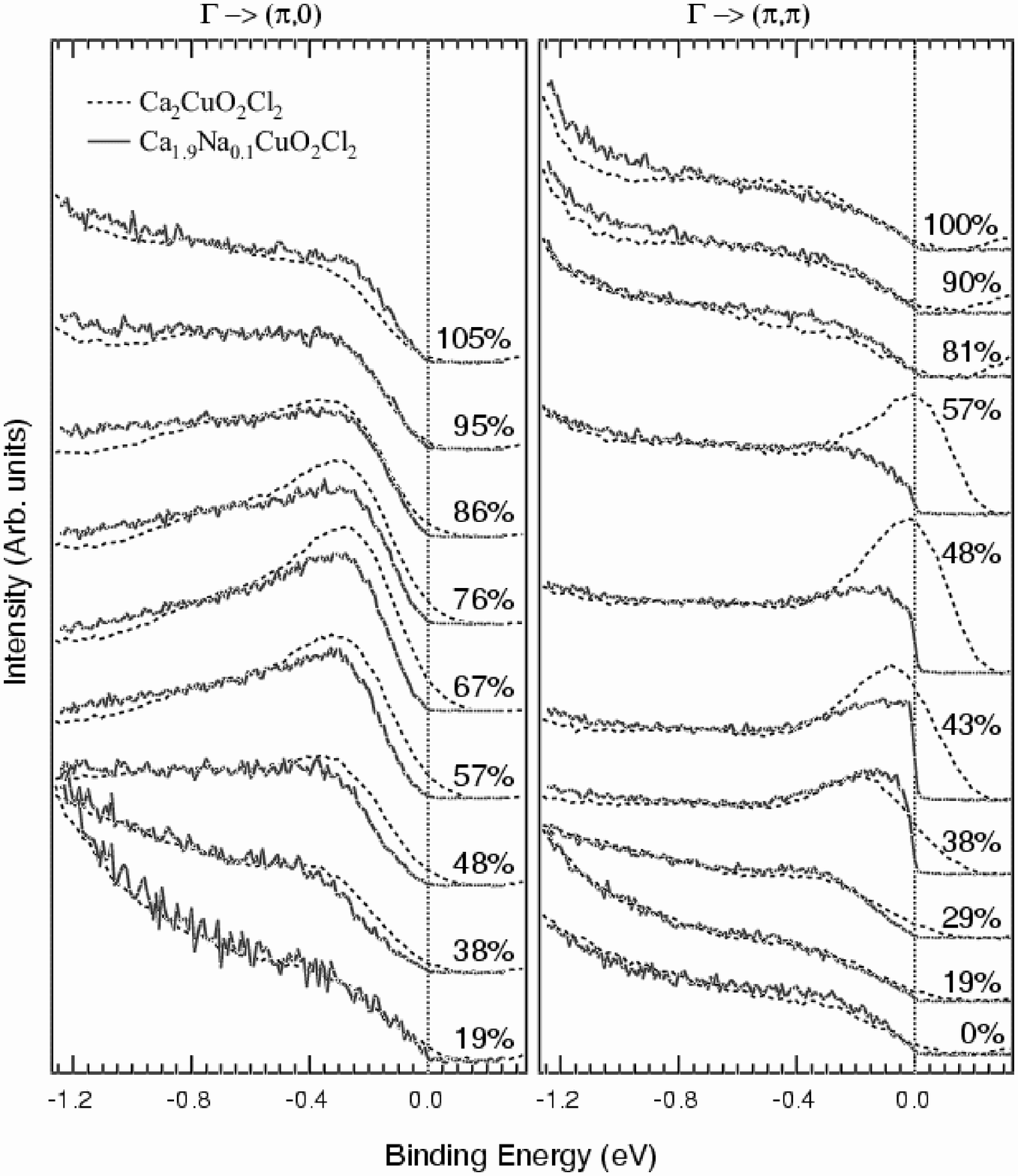} \vspace{0.1in}
\caption[EDC comparison 
of the metal to the insulator]{EDCs of 
Ca$_{2}$CuO$_{2}$Cl$_{2}$ (dashed line) (from ref 
\cite{RonningScience98}) are shifted by a constant in energy and 
compared with EDCs of 10$\%$ Na-doped CCOC(solid line).  The overlap is
extremely good with the exception of the features near 
$(\pi/2,\pi/2)$.  The data are normalized at high binding energy for 
comparison.  The slight rise in spectral weight above E$_{F}$ observed in 
some EDCs of CCOC is due to the presence of a core level excited by 
second order light.  This is not present in the metallic sample where 
the second order light contribution is heavily suppressed by the use 
of a normal incidence monochromator as opposed to the grazing incidence 
monochromator used for the insulating data. $E_{\gamma}$=25.5 eV, T=100 K 
and 20 K for x=0 and x=0.1, respectively.}
\label{FigNa3}
\end{figure}

To further support this idea, figure \ref{FigNa3} presents spectra near
E$_{F}$ along the entire (0,0) to $(\pi,0)$ and (0,0) to $(\pi,\pi)$
lines, where all the spectra from this insulating data set have been
shifted by 750 meV so that the centroid of the feature at $(\pi/2,\pi/2)$
is at the chemical potential. The value of 750 meV is specific to this
particular data set, but is 300 meV more than allowed by the uncertainty 
in
the position of the chemical potential.  Along (0,0) to $(\pi,0)$ the
spectra are remarkably similar.  In many instances fine details of the
line shape match extremely well.  For example, a second broad component in
the electronic structure can be observed at roughly 600 meV higher binding
energy from the first feature creating a heavily asymmetric
lineshape.\cite{KimPRL98,RonningThesis} This indicates that upon doping
the chemical potential simply drops to intersect the top of the valence
band similar to the case of an ordinary band material.  This would
naturally explain the large high energy pseudogap seen at $(\pi,0)$ as a
remnant property of the insulator as first suggested by
Laughlin.\cite{LaughlinPRL97,RonningScience98}

Along (0,0) to $(\pi,\pi)$ the situation is less clear.  As the feature
moves towards lower binding energy a sharp Fermi cutoff appears in the
metallic samples, as anticipated by the large peak at the Fermi level from
the energetically shifted spectra of the insulating sample.  However, the
match between the spectra becomes increasingly worse as $(\pi/2,\pi/2)$ is
approached.  Perhaps this simply indicates that a Fermi crossing has
occurred before $(\pi/2,\pi/2)$, and thus the weight at $(\pi/2,\pi/2)$ is
suppressed relative to the insulator or that some spectral weight transfer
in addition to the shift of the chemical potential has occured.  To see if
the suppression was due to matrix element effects we also changed the
polarization to maximize the nodal direction cross section, but it
remained weaker. This suppression possibly indicates that increased
scattering which could effectively smear out the $k$ resolution, and would
have the greatest effect where the dispersion is the steepest (near
$(\pi/2,\pi/2)$), has occured.  However, this would require a nontrivial
small angle scattering mechanism, which does not simply produce an angle
independent background.  Also, the cleavability of the Na-doped compounds
is the same as the parent insulator, and laser reflections from the sample
indicate flat surfaces, which argue against angle averaging as the cause
for the observed suppression of weight.  Typically, highly angle dependent
valence band spectra as shown in figure \ref{MvsIVB} are indicative of a
good surface, although the Na-doped valence band features are somewhat
broader than in the case of the insulator.  Pothuizen {\it et
al.}\cite{PothuizenPRL97} noted that in SCOC the lineshape of the 2 eV
binding energy peak at $(\pi,\pi)$ is identical to the lineshape of the
Zhang-Rice singlet at $(\pi/2,\pi/2)$. Given such a correlation a
broadened low-energy spectral lineshape at $(\pi/2,\pi/2)$ in Na-doped
CCOC is perhaps expected when the valence band at 2 eV has been broadened,
as observed here (see figure \ref{MvsIVB}). This suggests that the small
differences between the doped system and the half filled insulator may
indeed result from a difference in sample surface quality.

\begin{figure}[tb]
\centering 
\includegraphics[width=3in]{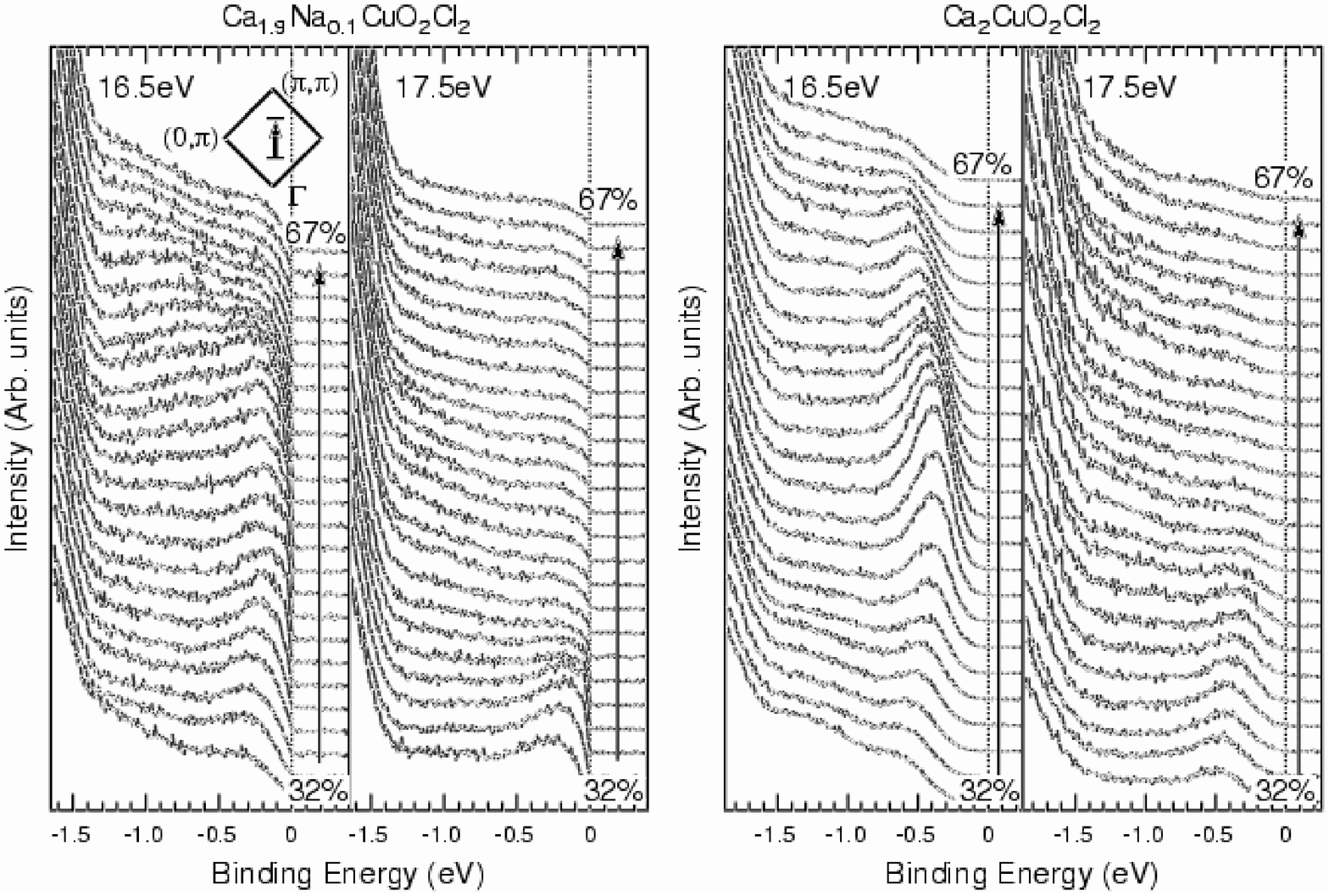}
\vspace{0.1in}
\caption[Photon energy dependence comparison of the metal to the 
insulator]{Photon energy dependence along the nodal 
direction for Ca$_{2-x}$Na$_{x}$CuO$_{2}$Cl$_{2}$ for x=0.1 and x=0.  
For each doping, the same $k$ range is shown, while the photon energy is 
indicated in each panel.  T=20 K and 293 K for x=0.1 and 0.0, 
respectively.  Notice, that the modulation of intensity varies with 
photon energy in a similar fashion for both dopings.}
\label{FigNa6}
\end{figure}

Considering that the features of hole doped CCOC track the dispersion of
the features in the half filled insulator, one might wonder exactly how
similar are the electronic states for the two doping levels.  A good way
to test this is by analyzing the photoemission matrix elements.  Under the
dipole approximation the photoemission intensity can be written as $I
\propto |\langle\Psi_{f}|A\cdot p|\Psi_{i}\rangle|^{2}$ where $\Psi_{i,f}$
are the initial and final state wavefunctions, and $A\cdot p$ is the
perturbing Hamiltonian.\cite{HufnerBook} Practically, this means that the
cross section of the photoemission process will depend on the photon
energy and experimental geometry as well as the wavefunctions themselves.  
Thus, if two wavefunctions are to be considered as the same, then they
should have the same dependence on the experimental conditions.  We have
found that CCOC experiences a fairly dramatic change in the modulation of
intensity along the nodal direction when changing the photon energy from
16.5 to 17.5 eV. In figure \ref{FigNa6}, we compare the EDCs along the
nodal direction of a 10$\%$ Na-doped CCOC sample and an undoped CCOC
sample using 16.5 eV and 17.5 eV photons.  We notice that the change in 
the
modulating intensity on going from $E_{\gamma}$=16.5 to 17.5 eV is similar
for both samples.  Namely, for 16.5 eV photons there is much more 
intensity
at $(\pi/2,\pi/2)$ and higher $k$ values, which vanishes on going to
17.5 eV photons.  Even the existence of the second feature at 600 meV 
higher
binding energy from the lowest energy feature can be seen in both samples
along the nodal direction.  These results show that the gross features of
the wavefunction of the electronic states within 2 eV of the chemical
potential discussed thus far are remarkably similar. The peculiar photon
energy dependence shown in figure \ref{FigNa6} will be discussed
separately with a much larger body of data.\cite{kyleunpublished} Along
with the fairly rigid shift in dispersion of the low energy states and the
valence band, the lineshape comparison and photon energy dependence show
conclusively that the chemical potential indeed shifts upon doping the
half filled insulator, Ca$_{2}$CuO$_{2}$Cl$_{2}$, with Na.

\subsection{Metallic Nature of Na-doped Ca$_{2}$CuO$_{2}$Cl$_{2}$}

\begin{figure}[t!]
\centering
\includegraphics[width=3in]{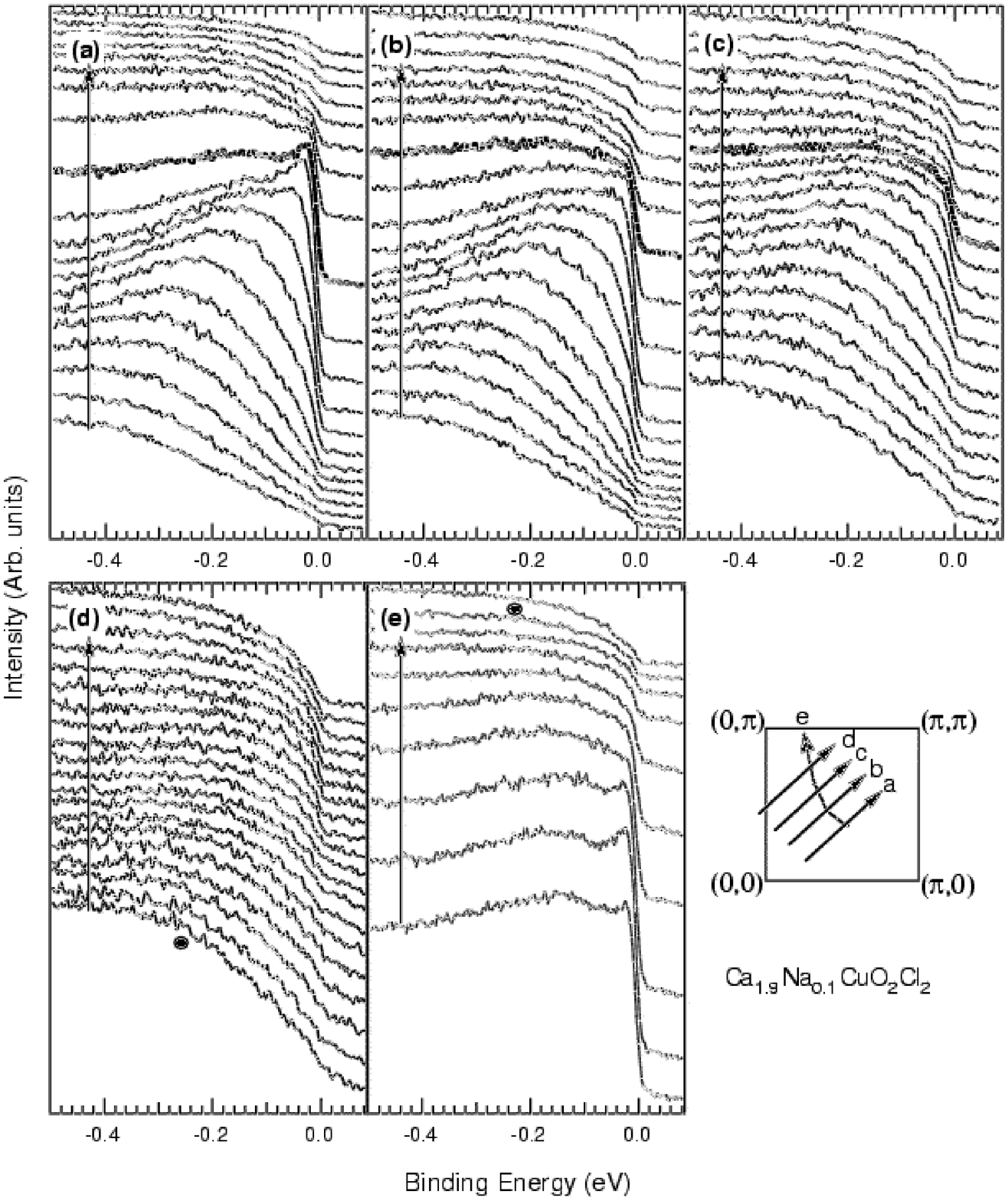}
\vspace{0.1in}
\caption[short caption1]
{Selected EDCs of Na-doped Ca$_{2}$CuO$_{2}$Cl$_{2}$ taken at the $k$ 
points
shown by solid lines in the cartoons. The lines in the Brillouin 
zone indicate all the 
$k$ points where spectra were taken. The bold spectra in (a) through (c)
indicate $k_{F}$. No spectra are highlighted in (d) as no Fermi crossing 
is observed. Spectra along the Fermi surface predicted by 
band theory are shown in (e). The dots in (d) and (e) indicate a high 
energy feature as discussed in the text. $E_{\gamma}$=He I$\alpha$=21.2 
eV; 
T=22 K} \label{FigNature1}
\end{figure}

A naive expectation of a simple chemical potential shift upon doping is
the creation of a small hole pocket centered at $(\pi/2,\pi/2)$.  However
this contrasts with the majority of ARPES results from the optimal and
overdoped regime which find a large Fermi surface centered about
$(\pi,\pi)$ as predicted by band theory.\cite{DamascelliReview} We shall
thus take a closer look at the near E$_{F}$ electronic structure to see if
the simple picture of a chemical potential shift as illustrated in figure
\ref{mushiftcartoon}b is correct.  In other words, is the electronic
structure unchanged with the exception of the position of the chemical
potential? Figure \ref{FigNature1} presents a sampling of the EDCs at
selected $k$ points.  The cuts shown in (a) through (d) are taken parallel
to the nodal direction.  For several of the cuts close to the nodal
direction a broad feature starts at high binding energy and approaches the
Fermi energy with increasing $k$, where a sharp Fermi cutoff of the
spectra is indicative of a Fermi surface crossing. We will return to the
presence of a peak-dip-hump like structure near the nodal Fermi crossing
later. The Fermi crossings, which have been determined by the spectra with
the minimum leading edge midpoint (the binding energy where the spectral
intensity is half of the intensity at the peak position), have been
highlighted in the figure.  We note that while a minimum value of the
leading edge midpoint identifies the spectra which corresponds to $k_{F}$,
the value itself will be pushed back to higher binding energy if this
portion of the Fermi surface is gapped. As the cuts approach $(\pi,0)$ the
spectral features become less pronounced and it quickly becomes impossible
to identify a Fermi crossing as one can no longer clearly identify a peak
position to meaningfully extract a leading edge. This is best illustrated
in panel (e) where spectra along the ostensible Fermi surface predicted by
band theory are presented. Near the node, there is a clear Fermi cutoff at
E$_{F}$, but this feature disappears quickly as one approaches the
antinodal region.

\begin{figure}[t!]
\centering
\includegraphics[width=3in]{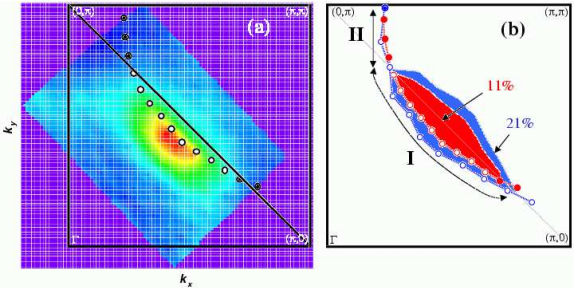}
\vspace{0.1in}
\caption[short caption]
{(color) (a) E$_{F}$ map obtained by integrating the x=0.10 spectra by 
$\pm$4 meV
about E$_{F}$. The open dots indicate the Fermi crossing, while the solid 
dots
extrapolate the Fermi surface to the Brillouin zone edge as would be
expected in band theory. (b) an illustration indicating the area of the
resulting Fermi surface pocket obtained by symmeterizing the observed
Fermi crossings about the antiferromagnetic zone boundary(dashed gray 
line). By counting
the number of electrons in the enclosed area a doping level is found
which is indicated in the figure. Red reproduces the points in (a) for
x=0.10, while blue are results from a sample with x=0.12. The data
naturally fall into two regimes: I(II) distinguishes the region along the
underlying LDA Fermi surface where a clear Fermi crossing does (does not)
occur.}
\label{FigIatEF}
\end{figure}

One can also visualize the data with an intensity map at E$_{F}$, obtained
by integrating each EDC by a $\pm$4 meV window about E$_{F}$. The highest
intensity contour, shown in figure \ref{FigIatEF}a, forms what appears as
a small arc.\cite{NormanNature98} The open circles indicate where a clear
Fermi crossing can be observed as in the cuts shown in figure
\ref{FigNature1}(a) through (c).  Note this ends as the Fermi surface
crosses the antiferromagnetic zone boundary.  By assuming that the Fermi
surface can not abruptly end in the middle of the Brillouin zone, one is
left with two possibilities.  Either the Fermi surface still lies along
the one predicted by band theory and has been heavily gapped as one
approaches $(\pi,0)$ or the Fermi surface is a small hole pocket centered
on or near $(\pi/2,\pi/2)$. We investigate these two possibilities 
below.

Let us first attempt to push the assumption that the large LDA-like Fermi
surface holds for this heavily underdoped sample. We separate the large
Fermi surface into two regions, as illustrated in figure \ref{FigIatEF}b.  
In Region I the broad spectral function has a sharp Fermi cutoff, which
clearly defines $k_{F}$, while in region II one must extrapolate the Fermi
surface from the Fermi crossings found in region I assuming a large Fermi
surface that does not terminate in the middle of the Brillouin zone.  
This is shown by the solid dots in figure \ref{FigIatEF}. Note that in the
region near $(\pi,0)$, variations in the precise shape of the Fermi
surface will not affect the following analysis as the spectra have little
variation which is apparent from the EDCs in figure \ref{FigNature1}d.

\begin{figure}[t!]
\centering
\includegraphics[width=3in]{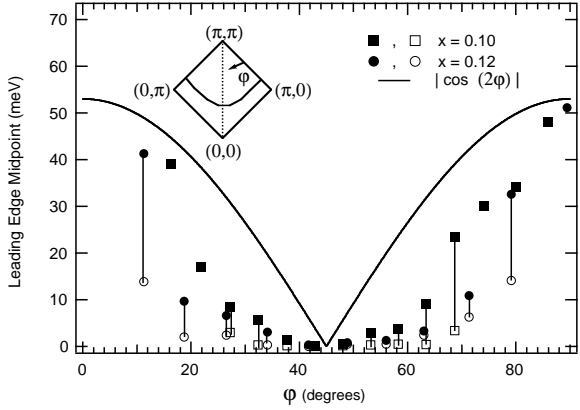}
\vspace{0.1in}
\caption[caption label]
{The leading edge midpoint of the spectra along the Fermi
surface illustrated in figure \ref{FigIatEF} relative to the 
value at the {\it d}-wave nodal position.  The open symbols are 
the result when
a clear Fermi cutoff can be identified, while the solid symbols are
obtained by normalizing the maximum of the low energy spectra to 1, as 
discussed in the text. The solid line is a plot of cos(2$\varphi$), the 
{\it d}-wave functional form.  The large discrepancy from the canonical 
{\it d}-wave functional form when attempting to describe the data in terms 
of a band theory-like Fermi surface centered at $(\pi,\pi)$
suggests that it is better to understand the data in terms of two
regimes, labeled I and II in figure \ref{FigIatEF}.}
\label{Figdwavedeviation}
\end{figure}

Figure \ref{Figdwavedeviation} plots the leading edge midpoint for spectra
along the large Fermi surface just discussed relative to the value at
$\phi$=45$^{\circ}$ (the nodal position for a pure d$_{x^{2}-y^{2}}$
functional form).  In this way one can determine the gap structure along
the Fermi surface. The open symbols indicate the leading edge midpoint for
the Fermi crossing spectra found in region I.  However, in region II the
lack of a clear leading edge midpoint prevented us from identifying a
Fermi crossing in the first place. Hence, we need an alternative method to
describe the dispersion.  Near $(\pi,0)$ an energy scale (albeit a broad
one) still exists in the spectra as indicated by a dot in figures
\ref{FigNature1} (d) and (e).  Thus we can blindly normalize the maximum
of the spectra below 300 meV to 1, and extract a leading edge midpoint via
this simplified approach.  This is what is shown by the filled symbols in
figure \ref{Figdwavedeviation}. For a simple spectral function this
procedure would give the correct gap value. In our case, one can see that
close to the node this procedure roughly matches our more careful
analysis, but it also allows us to extend our gap analysis beyond the
regime where a clear Fermi crossing has been identified, albeit in a crude
way. The solid black line is the expectation for the canonical {\it
d}-wave functional form. As mentioned above, in the optimal and overdoped
Bi2212 samples a simple spectral function and a large Fermi surface is
observed, and the {\it d}-wave functional form well describes the
experimental gap structure.\cite{ZXShenPRL93} However, in the heavily
underdoped regime studied here, there is a strong deviation from the {\it
d}-wave functional form. In order to reconcile this gap structure with
{\it d}-wave superconductivity\cite{footnote1} our assumption of a large
Fermi surface centered at $(\pi,\pi)$ must break down. In this case,
figure \ref{Figdwavedeviation} is not physically significant as it maps
the energy along two physically different regions as defined in figure
\ref{FigIatEF}b, each with its own gap structure. In other words, by
assuming a {\it d}-wave gap structure we have shown that only a fraction
of the spectra used for figure \ref{Figdwavedeviation} actually lie on the
Fermi surface.

Region I does indeed represent a portion of the Fermi surface about
$(\pi/2,\pi/2)$ which may contain a very small gap, while region II does
not contain a Fermi crossing, and the resulting large gap identifies a
larger energy scale.  We have just ruled out the possibility that the
Fermi surface in this sample is a large hole pocket centered at
$(\pi,\pi)$. If we maintain that the Fermi surface can not abruptly end in
the middle of the Brillouin zone we must conclude that the Fermi surface
forms a small hole pocket, as naively expected from the observed chemical
potential shift. The question then becomes, how does such a Fermi surface
develop a gap consistent with {\it d}-wave superconductivity. Given the
experimental resolution we can not comment on whether or not the flat
region in figure \ref{Figdwavedeviation} would itself fit a $|cos(k_x
a)-cos(k_y a)|$ functional form, which is the simplest expectation for
{\it d}-wave pairing, although models exist with small hole pockets and
{\it d}-wave pairing for which this is not the case.\cite{WenLeePRL98} We
note that for a superconductor with a transition temperature of 13 K, the
maximum of the gap from BCS theory is expected to be less than 2 meV at
the antinode and even smaller along the observed arc, which is too small
for our experimental resolution and is not the objective of this study.
Perhaps this gap could be measured with an improved spectrometer.

By symmeterizing the hole pocket about the antiferromagnetic zone boundary
one finds an area of 5.7$\%$, which implies a doping level of 11.4$\%$ due
to the electron spin. This is in rough agreement with the hole doping of
10$\%$ determined by comparing T$_c$ and lattice constants with results on
powdered samples.\cite{Hiroi} Again we note that the hole pocket is also
not necessarily symmetric about the antiferromagnetic zone
boundary,\cite{WenLeePRL98} which could substantially modify these values.
The idea of a small hole pocket about $(\pi/2,\pi/2)$ in Na-doped CCOC is
also consistent with the observations of shadow bands\cite{Kampf} at
higher binding energy.\cite{Kohsaka2001} Presumably, the coherence factors
responsible for the shadow band are very weak making it difficult to
distinguish a second Fermi crossing from the background at E$_{F}$. We
stress that while our findings are quite consistent with the small Fermi
pocket picture, we did not see the shadow Fermi surface at zero energy
even though we do observe a shadow band at higher energies.

Meanwhile, the observation of a chemical potential shift in this material
from the insulating state and the photon energy dependence of the
lineshape emphasize that the features near $(\pi,0)$ (region II) in the
Na-doped Ca$_{2}$CuO$_{2}$Cl$_{2}$ sample are a direct result of the
magnetism which dominates the antiferromagnetic insulator.  To illustrate
that the energy scale near $(\pi,0)$ is the same as that of the insulator,
recall figure \ref{FigNa3} where the spectra of the insulator matches that
of the Na-doped sample at $(\pi,0)$.

We have also repeated the above gap structure analysis on a sample with
x=0.12 (T$_c$=22 K), and some of the results are presented in blue in
figure \ref{FigIatEF} and as circles in figure \ref{Figdwavedeviation}.  
The data reproduce the above analysis with a few small quantitative
differences.  The Fermi arc increases in size (shown in figure
\ref{FigIatEF}b), consistent with an increase in doping, while the binding
energy near $(\pi,0)$ reduces slightly which demonstrates that this energy
scale is decreasing with increasing doping.  This is also expected as we
have identified this energy scale with the magnetism of the insulator
whose strength should diminish together with the antiferromagnetic
correlations as holes are added to the system.  This has also been
observed in previous studies on the high energy
pseudogap.\cite{WhitePRB96} We should note that the hidden message of
describing the Fermi surface in the underdoped regime as a small pocket
about $(\pi/2,\pi/2)$ is that the system must at some point transform upon
doping to the large Fermi surface centered about $(\pi,\pi)$ observed in
the overdoped regime.  In measuring the volume of the pocket for the
x=0.12 sample we find the doping level to be 21$\%$. The fact that this
analysis begins to show a deviation from a doping level of x perhaps
begins to identify a transition to a state where the doping level should
be described as 1-x (the expected volume of the large Fermi surface
centered at $(\pi,\pi)$).\cite{UchidaPRB91}

\begin{figure}[tb]
\centering 
\includegraphics[width=3in]{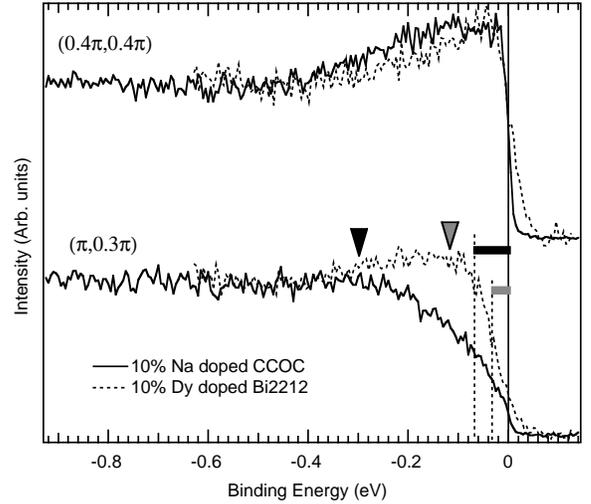}
\vspace{0.1in}
\caption[Comparison of EDCs between 10$\%$ Dy-doped Bi2212 and 10$\%$ 
Na-doped Ca$_{2}$CuO$_{2}$Cl$_{2}$]
{Comparison of EDCs at selected $k$ points of 10$\%$ 
Dy-doped Bi2212 (taken from ref \cite{MarshallPRL96}) and 10$\%$ 
Na-doped CCOC.  The Na-doped CCOC system 
suggests a much larger pseudogap than in Bi2212, both for the low 
energy pseudogap (thick bars) and the high energy pseudogap 
(triangles).}
\label{FigNa2}
\end{figure}

In comparison with other cuprates, the lack of a Fermi cutoff in the EDC's
would normally be identified as an extremely large pseudogap.  In an
attempt to quantify this suppression of weight, we compare 10$\%$ Na-doped
CCOC to 10$\%$ Dy-doped Bi2212\cite{MarshallPRL96} in figure 
\ref{FigNa2}.  
10$\%$ Dy-doped Bi2212 is underdoped with a T$_{c}$ of 65 K giving it a
T$_{c}$/Max(T$_{c}$) of 0.72, while 10$\%$ Na-doped CCOC has a ratio of
0.46. A comparison with 17.5$\%$ Dy-doped Bi2212 (not shown) with a
T$_{c}$/Max(T$_{c}$) ratio of 0.28 yielded a similar comparison. One
clearly observes the loss of spectral weight at the Fermi energy for the
Fermi momentum of $(\pi,0.3\pi)$ when compared to the respective Fermi
crossings along the nodal direction near $(0.4\pi,0.4\pi)$.  There are two
common methods for characterizing the pseudogap with ARPES. The first, is
by the shift of the leading edge midpoint of a spectra to higher binding
energy relative to the chemical potential as indicated by the dashed
lines.  This is typically referred to as the low energy pseudogap.  For
Dy-doped Bi2212 this is reasonably well defined, and gives a value of 30
meV. For the Na-doped CCOC sample, we find a value of roughly 50 meV at
$(\pi,0.3\pi)$.  However, as previously noted, the lack of a Fermi cutoff
near $(\pi,0)$ implies that the low-energy pseudogap is not well defined
in the Na-doped CCOC spectra. As a result, we are inclined to characterize
the pseudogap by what is known as the high energy pseudogap which
identifies a larger energy scale in the spectra about $(\pi,0)$, and is
sometimes referred to as a ``hump''.\cite{WhitePRB96} Here, we find a high
energy pseudogap of roughly 300 meV for Na-doped CCOC, compared with 
120 meV
for Dy-doped Bi2212 as indicated by the triangles in figure \ref{FigNa2}.  
Independent of the method chosen, the pseudogap for Na-doped CCOC appears
significantly larger than that of Bi2212 at a comparable doping level.

The high energy pseudogap seen in Bi2212 was first conjectured by Laughlin
to be a result of the $d$-wave-like modulation of the dispersion of the
insulating oxy-halides.\cite{LaughlinPRL97,RonningScience98} From the
observed shift of the chemical potential it is clear that the high energy
pseudogap indeed directly results from the dispersion seen in the parent
compound, although it appears that the energy scales of the high energy
pseudogap have a small system dependent variation.  The implication for
such a large pseudogap relative to the Bi2212 system at comparable doping
should be investigated, and the bilayer effect in Bi2212 will need to be
taken into consideration. If the condensation energy for superconductivity
is acquired by the gapping of low energy states near $(\pi,0)$, then this
would result in superconductivity being less favorable as evidenced by the
smaller optimal T$_{c}$(28 K and 90 K for Na-doped CCOC and Bi2212, 
respectively).

\subsection{Nodal Peak-Dip-Hump Structure}

A two component structure can be seen in the low energy spectra of
Na-doped CCOC at the Fermi crossing in the nodal direction.  This was
shown in figure \ref{FigNature1}e, where we plotted EDCs along the Fermi
arc and then towards $(\pi,0)$ as indicated in the cartoon inset.  The
reason for the difference in clarity of this structure between cleaves is
unknown, but is likely due to a variation of inhomogeneities and
experimental conditions which make it impossible to resolve the two
components that appear as one in the $(0.43\pi,0.43\pi)$ spectra of figure
\ref{FigNa3}.  Note that in general, the presence of sharp low energy
features is a tribute to a high sample quality.  The loss of the two
component structure as one approaches $(\pi,0)$ is indicative of the
observed large pseudogap, which precludes the identification of any
similar structure in the low energy excitations near $(\pi,0)$.

\begin{figure}[tbp]
	\centering
	\vspace{0.1in}
	\includegraphics[width=3in]{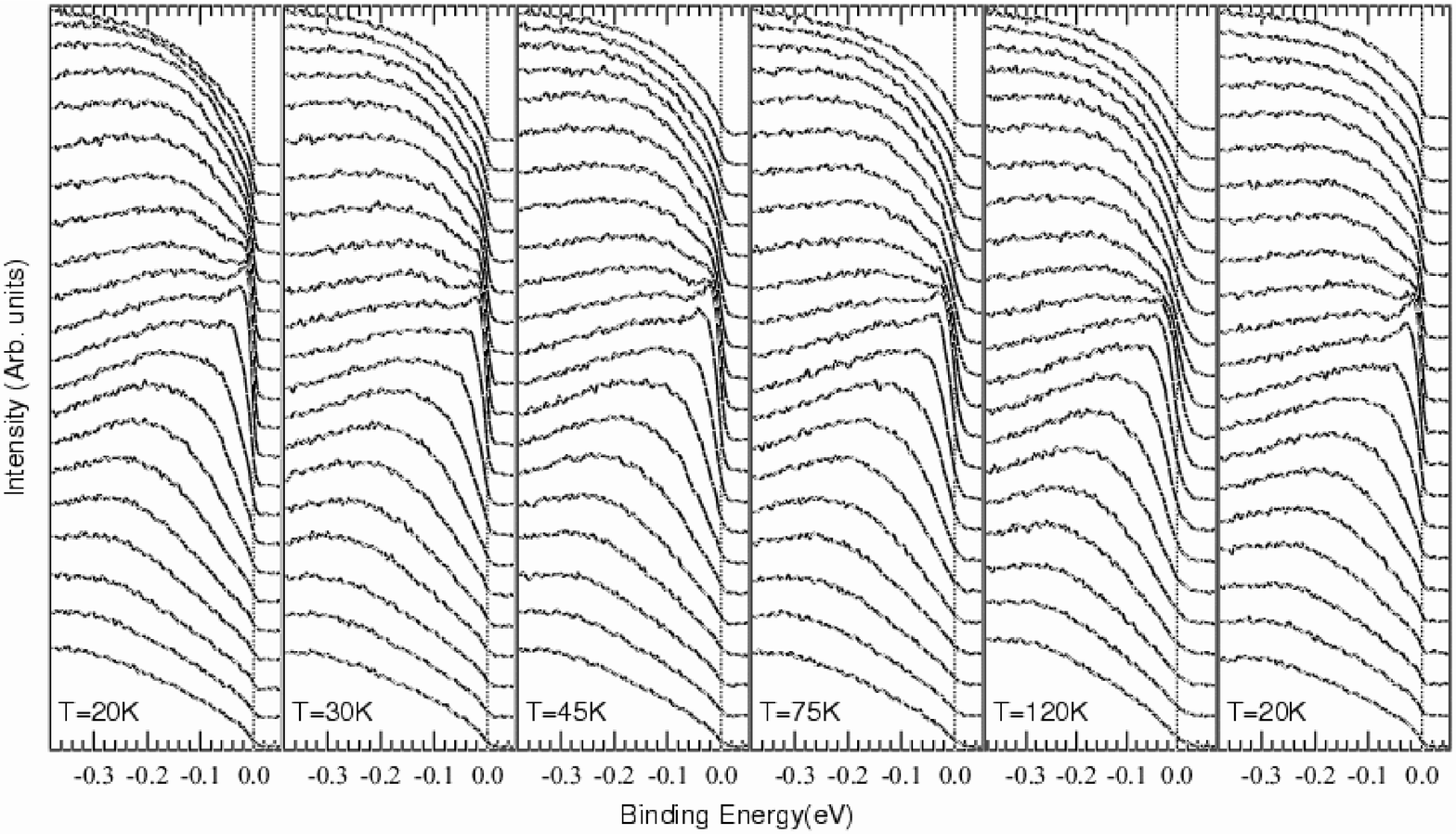}
	\vspace{0.1in}
\caption{a) Temperature dependence of 12$\%$ Na-doped CCOC 
along the nodal direction. A two component structure is observed at 
$k_{F}$ well above T$_{c}$. $E_{\gamma}$=16.5 eV}
	\label{peakdiphump}
\end{figure}

The observation of a two component structure in these superconducting
samples begs the question of how it may be related to superconductivity.  
Figure \ref{peakdiphump} presents temperature dependence of this feature
from 20 K($\approx$ T$_{c}$) up to 120 K. It is clear that the
peak-dip-hump like structure is observed up to 75 K, and is finally
smeared out at 120 K. By cycling the temperature back down to 20 K, we
ensure that the broadening at high temperatures is not an aging effect.  
Since even optimally doped polycrystaline samples have only a maximum
T$_{c}$ of 28 K, it is clear that the two component structure observed
here does not turn on near T$_{c}$.  This is in contrast with the more
familiar two component structure in ARPES on the Bi2212 system which has
now also been seen in YBCO\cite{LuPRL01} and Bi2223,\cite{FengBi2223} and
does turn on with T$_{c}$.\cite{FengScience00,Ding???} It is important to
note that, aside from the different temperature dependence, the two
component structure observed in Na-doped CCOC occurs along the nodal
direction as opposed to the $(\pi,0)$ region as in the other cases.
Furthermore, we note that CCOC is a single layer material and thus the
peak-dip-hump seen here can not be a consequence of bilayer splitting as
has been conjectured for the peak-dip-hump structure observed in 
Bi2212.\cite{KordyukPRL2002}

\begin{figure}[tbp]
\centering 
\includegraphics[width=3in]{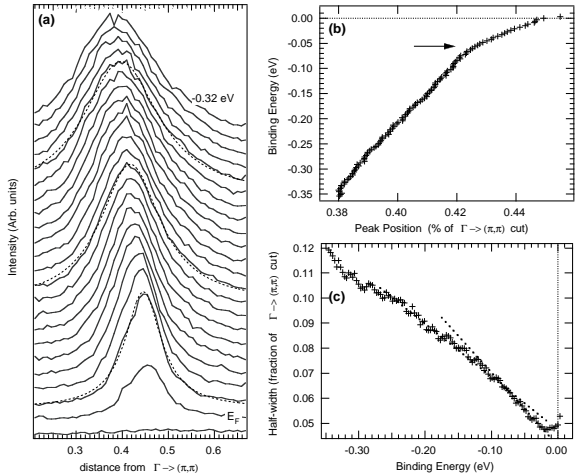} 
\vspace{0.1in}
\caption[MDC 
analysis of Ca$_{1.9}$Na$_{0.1}$CuO$_{2}$Cl$_{2}$]
{a) Sample MDCs of Ca$_{1.9}$Na$_{0.1}$CuO$_{2}$Cl$_{2}$ from the nodal 
direction taken with He I$\alpha$ 
radiation.  Single Lorentzian fits to selected MDCs 
are overlayed with a dotted line.  Respectively, b) and c) give the peak 
positions 
and widths (half width at half maximum) from the MDC analysis.  The 
arrow in b) indicates the kink energy, as discussed in the text.}
\label{FigNa11}
\end{figure}

In order to gain some insight into the origin of this structure figure
\ref{FigNa11} presents a MDC analysis along the nodal direction of a
x=0.10 sample.  The dashed curves are the single Lorentzian fits to
selected MDCs.  Similar procedure in other cuprates has been used to
extract self energies from the fit
parameters.\cite{LanzaraNature,VallaPRL00,kinkfound} Here the fits serve
as a method for parameterizing the observed dispersion.  Figures
\ref{FigNa11}b and \ref{FigNa11}c plot the peak positions and widths as a
function of the binding energy of the corresponding MDCs.  A change of
slope in the dispersion of the peak positions is seen near 55 meV. This
value is attained from the intersection of two linear fits ranging from 0
to 30 meV and 100 to 200 meV. We shall refer to this feature as a 
``kink''.  
A similar kink in other cuprate systems has received significant attention
in recent ARPES
literature.\cite{LanzaraNature,kinkfound,KaminskiPRL01,Johnsonxxx01}
Certainly, the similarity of the kink energy position to the dip energy
position seen in the EDCs suggests that these two results are related.  
The MDC width also displays an accelerated decrease in width as indicated
by the dotted lines in figure \ref{FigNa11}c.

What is the possible origin of a two component structure in the EDC near
$k_{F}$ along the nodal direction?  Pure macroscopic phase separation of
metallic and undoped insulating domains can be ruled out, since the
feature seen in the insulator would not have shifted to the chemical
potential in this case.  One possibility is that a sharp, coherent-like
peak is growing with doping, while the incoherent features at half filling
slowly vanish. This idea is captured in the recent phase string
calculations done by Muthukumar, Weng, and Sheng.\cite{PhaseString} In
this sense, the two components would hint at a balance between the
antiferromagnetic insulator and the drive for the system to become
metallic.  An alternative scenario is that the peak-dip-hump structure is
the result of coupling to a collective mode.  This could also naturally
explain the presence of the kink. Although still a debated interpretation,
Lanzara {\it et al.} propose that the universally observed kink is due to
electron-phonon coupling.\cite{LanzaraNature} In this regard, it is
interesting to note that the phonon breathing mode is expected to be
roughly 10 meV less for CCOC on the basis of its a-axis lattice constant
than in the case of insulating Bi2212 or LSCO as determined by
optics,\cite{TajimaPRB91} while the kink energy position determined by
ARPES is less by roughly the same amount.

Distinguishing between these two scenarios will be difficult.  A problem
with the first picture would be that the kink would have to be explained
as an artifact of the MDC analysis due to the multiple electronic features
which are not individually resolved.  This is not appealing considering
the apparent universality of the kink feature among
cuprates,\cite{LanzaraNature} and one would then favor a coupling to a
collective mode.  However, it is surprising that the kink is seen in both
LSCO where the states were created inside the gap, and in Na-doped CCOC
where the states appear to originate from the effective lower Hubbard band
which has now been shifted to the chemical potential.  This suggests that
a third possibility, combining both ideas, may also be considered for the
doping evolution.  Namely, that the chemical potential has shifted upon
doping, the precise location would be system dependent, while states are
created at E$_{F}$ which then couple to some collective excitations
creating the universally observed kink (see figure \ref{mushiftcartoon}d).
If this is the case, the Ca$_{2-x}$Na$_{x}$CuO$_{2}$Cl$_{2}$ and
La$_{2-x}$Sr$_{x}$CuO$_{4}$ system would no longer appear so different.  
Unfortunately, the present data can not distinguish between these three
possibilities, and we must leave this as an opportunity for future
experiments with more available doping levels to investigate.

\section{Conclusions and Discussion}

The data from the valence band, near E$_{F}$ spectra, and photon energy
dependence in the Ca$_{2-x}$Na$_{x}$CuO$_{2}$Cl$_{2}$ system provide
convincing evidence that the chemical potential shifts to the top of the
valence band upon doping the insulator, Ca$_{2}$CuO$_{2}$Cl$_{2}$.  
Surprisingly, these results appear to present a different evolution across
the metal to insulator transition than in LSCO where the chemical
potential is pinned and states are created inside the gap upon
doping.\cite{InoPRB00} 

With the comparison of CCOC we have shown that the high energy pseudogap
is indeed a remnant property of the insulator.\cite{LaughlinPRL97}
Meanwhile, by examining the low lying excitations we found that an attempt
to use a large Fermi surface centered at $(\pi,\pi)$ results in a gap
structure which deviates from the canonical {\it d}-wave functional form.  
Instead a small Fermi pocket about $(\pi/2,\pi/2)$ as naively expected
from a rigid chemical potential shift is a better description of heavily
underdoped Ca$_{2-x}$Na$_{x}$CuO$_{2}$Cl$_{2}$.  This is the case when we
insist that the Fermi surface can not terminate at a point in the middle
of the Brillouin zone. A Fermi surface arc could also explain the data if
this were allowable in a non-Fermi liquid picture. We anticipate this
region to also possess a gap structure which could differentiate between
competing models in the underdoped regime. Finally, a two component
structure is also observed in the EDCs along the nodal direction. This
appears to be related to a kink seen in the MDC derived dispersion, which
is now a universally observed feature of the cuprates.\cite{LanzaraNature}

The presence of features seen in the insulator within the low energy
spectra of the Na-doped superconductor suggests a strong interplay between
antiferromagnetism and superconductivity in the underdoped regime.  One
way to think of these results is to start from the overdoped regime where
superconductivity is the first instability encountered.  Upon further
underdoping, magnetism begins to play a role which opens a large
pseudogap, and suppresses superconductivity.  Alternatively, one can start
with the insulator, which forms a small pocket as carriers are introduced
into the system. Superconductivity then creates a new instability that
opens a small gap along the Fermi surface, which we are unable to detect
given our resolution. We look forward to future experiments which will
clarify Ca$_{2-x}$Na$_{x}$CuO$_{2}$Cl$_{2}$'s
evolution to the more "conventional" behaviour of the overdoped regime,
and its relation to the other cuprate superconductors.

\section{Acknowledgments}

F.R. would like to thank M.B. Walker and N. Schwanger for generously
providing assitance to fascilitate completion of this work, as well as 
A.-M.S. Tremblay and S.C. Zhang for helpful discussions. This research
was carried out at the Stanford Synchrotron Radiation Laboratory which is
operated by the D.O.E. Office of Basic Energy Science, Division of
Chemical Sciences. The Office's Division of Material Science provided
funding for this research. The Stanford work is also supported by NSF
grant DMR-0071897.


\begin{thebibliography}{99}
 
\bibitem[\ast]{fr}  {\it Present Address}: Dept. of Physics,
University of
Toronto, 60 St. George St, Toronto, ON, M5S 1A7, Canada
 
\bibitem[\dag]{dlf}  {\it Present Address}: Dept. of Physics and
Astronomy,
University of British Columbia, Vancouver, Canada
 
\bibitem[\ddag]{cyk}  {\it Present Address}: Dept. of Physics, Yonsei
University, Seoul, Korea
 
\bibitem[\S]{npa}  {\it Present Address}: Dept. of Physics and Astronomy,
UCLA, Los Angeles, CA
 
\medskip
\bibitem{TimuskRPP99} T. Timusk and B. Statt, Rep. Prog. Phys. {\bf 62},
61 (1999)
\bibitem{AndersonScience87} P.W. Anderson, Science {\bf 235}, 1196
(1987) 
\bibitem{MeindersPRB93} M.B.J. Meinders, H. Eskes, and G.A. Sawatzky, 
Phys. Rev. B {\bf 48}, 3916 (1993)
\bibitem{AllenPRB90} J.W. Allen {\it et al.}, Phys. Rev. Lett. {\bf 64},
595 (1990)
\bibitem{InoPRB00} A. Ino {\it et al.}, Phys. Rev. B {\bf 62} 4137
(2000) 
\bibitem{RombergPRB90} H. Romberg {\it et al.}, Phys. Rev. B {\bf 42},
8768 (1990)
\bibitem{ChenPRL91} C.T. Chen {\it et al.}, Phys. Rev. Lett. {\bf 66}, 104
(1991)
\bibitem{VanVeenendalPRB94} M.A. van Veenendaal, G.A. Sawatzky, and W.A.
Groen, Phys. Rev. B {\bf 49}, 1407 (1994)
\bibitem{HarimaPRB01} N. Harima {\it et al.}, Phys. Rev. B {\bf 64},
220507(R) (2001) 
\bibitem{Kohsaka2001} Y. Kohsaka {\it et al.}, To be published 
\bibitem{LaughlinPRL97} R.B. Laughlin, Phys. Rev. Lett. {\bf 79}, 1726
(1997) 
\bibitem{RonningScience98} F. Ronning {\it et al.}, Science {\bf 282}, 
2067
(1998).
\bibitem{LanzaraNature} A. Lanzara {\it et al.}, Nature {\bf 412} 510
(2001) 
\bibitem{KohsakaGrowth} Y. Kohsaka {\it et al.}, J. Am. Chem. Soc. {\bf 
124} 12275 (2002)
\bibitem{Hiroi} Z. Hiroi, N. Kobayashi, and M. Takano, Nature, {\bf 371},
139 (1994); Z. Hiroi, N. Kobayashi, and M. Takano, Physica C, {\bf 266},
191(1996)
\bibitem{BeardSSS94} B.C. Beard, Surf. Sci. Spectra {\bf 2}, 91
(1994); Surf. Sci. Spectra {\bf 2}, 97 (1994).
\bibitem{DurrPRB00} C. D\"urr, {\it et al.}, Phys. Rev. B {\bf 63},
014505 (2000)
\bibitem{PothuizenPRL97} J.J.M. Pothuizen {\it et al.}, Phys. Rev.
Lett. {\bf 78}, 717, (1997)
\bibitem{KimPRL98} C. Kim {\it et al.}, Phys. Rev. Lett. {\bf 80}, 4245,
(1998)
\bibitem{RonningThesis} F. Ronning, PhD. thesis, Stanford University
(2001)
\bibitem{HufnerBook} S. H\"{u}fner, {\it Photoelectron spectroscopy:
principles and application}, New York: Springer-Verlag, c1995.
\bibitem{kyleunpublished} K.M. Shen, et al. To be published.
\bibitem{DamascelliReview} A. Damascelli, Z. Hussein, and Z.-X. Shen,
submitted to Rev. Mod. Phys. (2001); D.W. Lynch and C.G. Olson, {\it
Photoemission Studies of High-Temperature Superconductors} Cambridge
University Press, Cambridge (1999); Z.-X. Shen and D. S. Dessau, Phys.
Rep. {\bf 253}, 1 (1995) and references therein.
\bibitem{NormanNature98} M.R. Norman {\it et al.}, Nature {\bf 392}, 157
(1998) 
\bibitem{ZXShenPRL93} Z.-X. Shen {\it et al.}, Phys. Rev. Lett. {\bf
70}, 1553 (1993)
\bibitem{footnote1} The Na-doped CCOC samples have been measured at a
temperature above T$_c$. However, ARPES work on underdoped Bi2212 find a
large Fermi surface with an identical gap structure above and below T$_c$
consistent with the {\it d}-wave functional form. See A.G. Loeser {\it
et al.}, Science {\bf 273}, 325 (1996); H. Ding {\it et al.}, Nature,
{\bf 382}, 51 (1996) 
\bibitem{Kampf} A.P. Kampf and J.R. Schrieffer,
Phys. Rev. B {\bf 42}, 7967, (1990) 
\bibitem{WenLeePRL98} See for example X.-G. Wen and P.A. Lee, Phys. Rev.
Lett. {\bf 80},
2193, (1998)
\bibitem{WhitePRB96} P.J. White {\it et al.}, Phys. Rev. B {\bf 54},
R15669 (1996)
\bibitem{UchidaPRB91} S. Uchida {\it et al.}, Phys. Rev. B {\bf 43},
7942 (1991) 
\bibitem{MarshallPRL96} D.S. Marshall {\it et al.}, Phys. Rev. Lett. {\bf
76} 4841 (1996)
\bibitem{LuPRL01} D.H. Lu {\it et al.} Phys. Rev. Lett. {\bf 86},
4370 (2001) 
\bibitem{FengBi2223} D.L. Feng {\it et al.} Phys. Rev. Lett. {\bf 88}, 
107001 (2002)
\bibitem{FengScience00} D.L. Feng {\it et al.}, Science, {\bf 280} 277
(2000)
\bibitem{Ding???} H. Ding {\it et al.}, Phys. Rev. Lett. {\bf 87}, 227001 
(2001)
\bibitem{KordyukPRL2002} A.A. Kordyuk {\it et al.} Phys. Rev. Lett. {\bf 
89}, 077003 (2002) 
\bibitem{VallaPRL00} T. Valla {\it et al.}, Phys. Rev. Lett. {\bf 85}
828 (2000)
\bibitem{kinkfound} P. Bogdanov {\it et al.}, Phys. Rev. Lett. {\bf 85}
2581 (2000)
\bibitem{KaminskiPRL01} A. Kaminski {\it et al.}, Phys. Rev. Lett. {\bf
86} 1070 (2001);
\bibitem{Johnsonxxx01} P.D. Johnson {\it et al.}, Phys. Rev. Lett. {\bf
87} 177007 (2001)
\bibitem{PhaseString} V.N. Muthukumar, Z.Y. Weng, and D.N. Sheng, Phys. 
Rev. B {\bf 65}, 214522 (2002) 
\bibitem{TajimaPRB91} S. Tajima {\it et al.}, Phys. Rev. B {\bf 43},
10496 (1991)

\end{thebibliography}
\end{document}